\documentstyle[psfig]{elsart}

\begin{document}
\begin{frontmatter} 
 
\title{
Separating True V0's from Combinatoric Background with a Neural Network}

\author{Marvin Justice\thanksref{email}}

\address{Department of Physics and Center for Nuclear Research, 
       Kent State University,
       Kent, OH 44242 }
\date{December 2, 1996}

\thanks[email]{Tel. +1 510 486 6531, e-mail justice@sseos.lbl.gov}

\begin{abstract}
 A feedforward multilayered neural network has been trained
 to ``recognize'' true V0's in the presence of a large
 combinatoric background using simulated data for 
 2 GeV/nucleon Ni + Cu interactions. The resulting neural
 network filter has been applied to actual data from the
 EOS TPC experiment. An enhancement of signal to  background
over more traditional selection mechanisms has been observed.
\end{abstract}
\end{frontmatter}

\section{Introduction}

A high statistics sample of $\Lambda$'s and K$^{0}_{\mathrm{s}}$'s
produced in 2 GeV/nucleon Ni + Cu
collisions has been obtained with the 
EOS Time Projection Chamber \cite{justice95}.
These neutral strange particles, or V0's,
are reconstructed through their charged
particle decays: $\Lambda \longrightarrow p + \pi^{-}$ and
K$^{0}_{\mathrm{s}} \longrightarrow \pi^{+} + \pi^{-}$ .
The acceptance plus efficiency for detecting true V0's
with the EOS TPC is very good; however, the sample
is contaminated by a large combinatoric
background of false $\Lambda$'s and K$^{0}_{\mathrm{s}}$'s
({\em e.g.\/}  $\sim40,000$ $p\pi^{-}$ pairs for every
true $\Lambda$). 

Traditionally, the signal is extracted from
the background by cutting on certain parameters ---
such as distance of decay from the main vertex --- whose
distributions are different for signal and background.
Inevitably, such cuts eliminate a significant fraction 
of the signal as well and one is confronted with the task
of how to optimize the cuts. The optimization problem
is a natural candidate for neural network techniques.
A feedforward, multilayered neural
network filter has been devised which, 
when applied to the EOS data, results
in a cleaner, higher statistics sample of $\Lambda$'s and K$^{0}_{\mathrm{s}}$'s.

\begin{figure}[ht]
\centerline{\psfig{figure=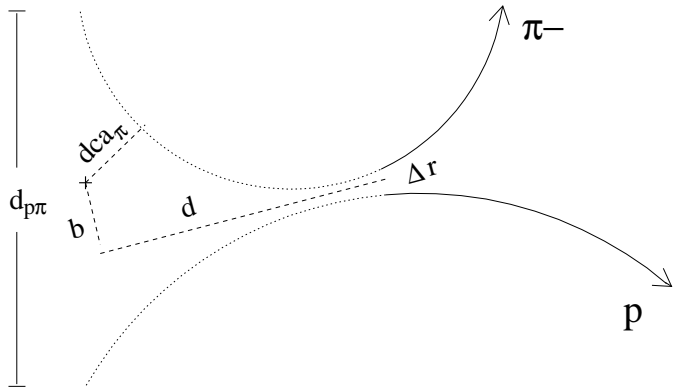,height=3in}}
 \caption {Schematic diagram illustrating $\Lambda$ decay and reconstruction.}
 \label{V0PAR}
\end{figure}

\section{V0 Reconstruction}

A schematic diagram illustrating $\Lambda$ reconstruction
and the set of parameters used to separate the signal
from the background is shown in Fig.~\ref{V0PAR}. 
V0 reconstruction begins after all
TPC tracks in an event have been 
found and the overall
event vertex has been determined.
In the case of $\Lambda$'s,
each pair of $p\pi^{-}$ tracks
is looped over and their point of
closest approach is calculated. Pairs
whose trajectories approximately intersect 
at a point other than the main vertex are fit with  a
V0 hypothesis from which the invariant mass,
momentum, and point of decay 
$(\vec{X_{\Lambda}})$ are extracted. 

The $\Lambda$ momentum vector is projected
back to the target to obtain $d$, the
distance between $\vec{X_{\Lambda}}$ and the overall
event vertex, and $b$, the impact parameter or
distance of closest approach between the 
event vertex and the $\Lambda$ trajectory.
Likewise, the proton and $\pi^{-}$ tracks
are projected back to the target to obtain
$dca_{p}$ and $dca_{\pi}$ which are the
distances of closest approach of the 
daughter particle trajectories to the main
vertex. Another useful variable, closely related to
the two $dca$'s, is the distance between the
$p$ and $\pi^{-}$ trajectories at the target
plane, $d_{p\pi}$; while the distance between the $p$ and $\pi^{-}$
trajectories at $\vec{X_{\Lambda}}$ is called
$\Delta r$. In an ideal detector $b$ and $\Delta r$ 
would be exactly zero for true $\Lambda$'s and K$^{0}_{\mathrm{s}}$'s.
In any real detector, of course, these quantities
will take on finite values. The 
estimated impact parameter resolution for the
EOS TPC is $\sigma_{b} \sim3$~mm.

Seven parameters are cut on to separate true
$\Lambda$'s and K$^{0}_{\mathrm{s}}$'s from 
the background:
$d$, $b$, $\Delta r$, $d_{p\pi}$, $dca_{p}$,
$dca_{\pi}$, and the $\chi^{2}/\nu$ of the
combined fit to the V0 hypothesis. The traditional
method is to simply define a seven 
dimensional cut such as:
$d \geq 4$~cm
AND $b \leq 3$~mm AND $d_{p\pi} \geq 2$~cm AND etc.
The problem then becomes what the precise values
of the cuts should be. For example, nearly all of
the combinatoric background can be eliminated by
simply requiring that $d$ be very large. Since
the true V0's follow an exponential decay law,
however, such a cut would throw out most of the true
signal as well. 

\begin{figure}[t]
\centerline{\psfig{figure=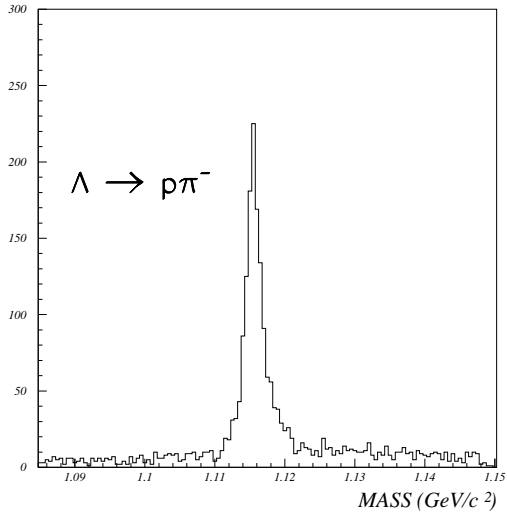,height=3in}}
 \caption  { $\Lambda$ invariant mass spectrum from seven parameter cuts.}
 \label{MLAM_TRAD}
\end{figure}

For the EOS data, an attempt to
optimize the cuts has been made through trial
and error using the invariant mass distribution
as a guide. The invariant mass distribution
resulting from the ``best'' cuts 
for $\Lambda$'s is shown in Fig.~\ref{MLAM_TRAD}. From Monte Carlo 
simulations it is estimated that over 60\% of 
the reconstructed true $\Lambda$'s are lost in making the cuts necessary
to obtain the background level in this plot.

Clearly, searching a seven dimensional parameter space 
by trial and error in an effort to optimize the cuts
can be a very tedious process. In addition, the high
degree of correlation among some of the parameters
makes it unlikely that the optimum cut would be 
obtained by cutting perpendicular to each of the
seven axes as above. 
An uncorrelated set of paremeters could be formed
from linear combinations of the original seven
parameters by using a principal component analysis.
However, one would still be left with the task of
optimizing the cuts in the new parameter space. 
Moreover, any boundary surface chosen to 
separate the true V0's from the background would 
still be restricted to a seven dimensional polyhedron.
An alternative method of
cutting, which may allow for more complicated
shapes in the seven dimensional space, is provided
by neural network techniques.

\section{Neural Network Approach}

\begin{figure}[t]
\centerline{\psfig{figure=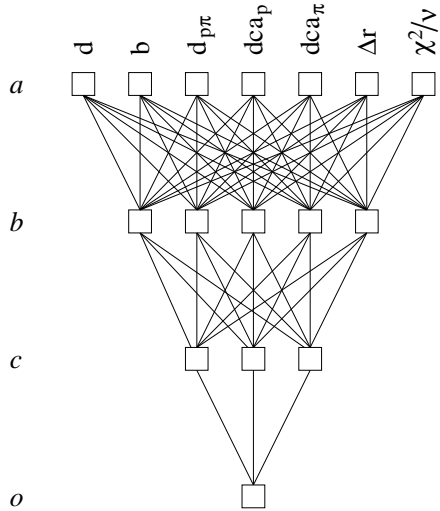,height=3in}}
 \caption { V0 neural network topology.}
 \label{TOPOLOGY}
\end{figure}

A general feedforward multilayered network consists
of a set of input neurons, one or more layers of 
hidden neurons, a set of output neurons, and 
synapses connecting each layer to the subsequent
layer \cite{mul91}. A  particular network topology  
for the appliction at hand 
is shown in Fig.~\ref{TOPOLOGY} where the inputs, $a_{i}$, are
the seven V0 parameters. There are two hidden layers, $b_{j}$
and $c_{k}$, and one output, $o$. The network is fully connected
in the sense that there is a synapse connecting each $a$ neuron
to each $b$ neuron, each $b$ neuron to each $c$ neuron, and
each $c$ neuron to the output. Given a set of inputs, the
rules for calculating $o$ are:
\begin{eqnarray}
b_{j} & = & \tanh\left(\sum_{i} w_{ij}^{ab} a_{i} - \Theta_{j}\right) \, , \\
c_{k} & = & \tanh\left(\sum_{j} w_{jk}^{bc} b_{j} - \Theta_{k}\right) \, , \\
o & = & \sum_{k} w_{k}^{co} c_{k} \, ,
\end{eqnarray}
where the $w$'s are synaptic weights and the $\Theta$'s
are thresholds. 

For the current application one would like $o$
to take on one value for true V0's and a different 
value for false V0's, {\em e.g.\/} +1 for true and --1 for false.
The problem then becomes one of finding a set of weights and
thresholds that give the desired outputs. In general,
this is accomplished by starting with random initial
guesses for the $w$'s and $\Theta$'s and ``teaching''
the network with a training set and a backpropagation
algorithm. 

A set of $\sim2.5 \times 10^{5}$ 2 GeV/nucleon Ni + 
Cu events generated with the ARC cascade 
code \cite{arc92} has been
used to train and test the network
of Fig.~\ref{TOPOLOGY}.
Only those events which had some
strangeness content in the final state
were used in the training stage.
The strange events were
run through a detailed GEANT simulation of the
TPC and passed through the same analysis
chain as was the real data. 
Loose cuts on the seven 
V0 parameters were applied to the output
in order to weed out easy background.
The resulting training set was composed of
3757 true V0's and 41,600 combinatoric V0's. 
The V0's were labeled 
as being either true or false based on 
information stored from GEANT
and  were then passed one at a time through the
neural net. Separate, though topologically identical, 
networks were used for
$\Lambda$'s and K$^{0}_{\mathrm{s}}$'s. 

As each V0
is passed through its appropriate network 
the weights and thresholds are adjusted
so that the actual  output approaches the desired output.
This is done by minimizing the error function:
\begin{eqnarray}
E & = & \frac{1}{2}(t - o)^{2} \nonumber \\
  & = & \frac{1}{2}(t - \sum_{k} w_{k}^{co} c_{k})^{2} \, ,
\end{eqnarray}
where $t = +1$ for true V0's  and
$t = -1$ for background or fake V0's.
A simple gradient
descent algorithm: 
\begin{eqnarray}
\Delta w_{ij}^{\alpha\beta} & = & 
                  -\eta\frac{\partial E}{\partial w_{ij}^{\alpha\beta}} \, ,\\
\Delta \Theta_{i}  & = & -\eta\frac{\partial E}{\partial \Theta_{i}} \, ,
\end{eqnarray}
is used to adjust the 
$w$'s and $\Theta$'s after $o$
is calculated for each V0.
In the present application $\eta$ was chosen to be 0.05 
and all of the thresholds were held fixed at zero.

In principle, the training process 
should continue until the weights cease 
to change. In practice the same set of 
events was passed repeatedly 
through the network in alternating training
and testing cycles. After each training cycle 
the events were filtered back through the network
and a histogram of the resulting outputs 
(similar to Fig.~\ref{NETOUT}) was
visually inspected to judge convergence.
The cpu time per training cycle was 
$\sim$12 minutes on a 55 MHz  HyperSparc.
After 10 cycles the networks were judged to have 
converged.

\section{Results}

\begin{figure}[t]
\centerline{\psfig{figure=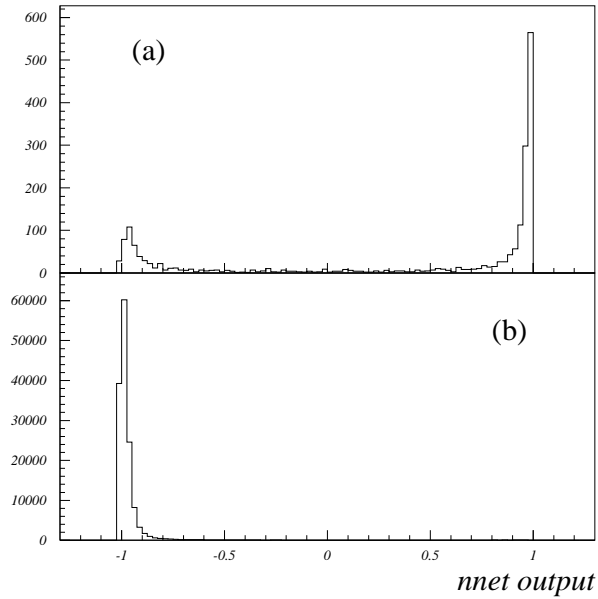,height=3.5in}}
 \caption { Neural network output for (a) true and (b) false
  Monte Carlo $\Lambda$'s.}
 \label{NETOUT}
\end{figure}

The performance of the networks was first tested
on a subset of $\sim1.73 \times 10^{5}$ ARC events which 
had not been preselected for strangeness. Although all of the 
events containing V0's in this subset were also 
members of the training set, they had been rerun 
through GEANT with different 
random number seeds. The same loose cuts
on the seven parameters were applied to the
GEANT output as in the training phase resulting in
2601 true and $\sim1.75 \times 10^{5}$  false V0's ---
roughly the same 1:7 ratio as is 
observed in the EOS data after loose cuts.
The V0's were passed
through the neural network filters and
the resulting distributions of ouputs for $\Lambda$'s
are shown in Fig.~\ref{NETOUT}.

Qualitatively, one sees that the $\Lambda$ network performs
as desired: the distribution for the true's has a sharp
peak at +1 while the combinatoric distribution is peaked
at --1. 
Quantitatively, one can define a purity:
\begin{displaymath}
Purity = \frac{true}{true + false} \,\, ,
\end{displaymath}
and a yield:
\begin{displaymath}
Yield = \frac{detected \,\, true}{actual \,\, true} \,\, .
\end{displaymath}
The purity and yield factors for the traditional method of
cutting and for cuts on the value of the neural
network output are listed in Table~1.
All yields reflect a common $\sim$60\% loss factor 
arising from
geometrical acceptance and tracking efficiency.
From the table  one sees that, for the Monte Carlo
events, the neural network method gives a significantly
higher yield than the traditional method at the same
level of purity. Alternatively, higher purity levels
can be obtained without loss of yield. Similar results
are obtained with the K$^{0}_{\mathrm{s}}$ network.

\begin{table}
\centering
\begin{tabular}{ccc} \hline
  cut method   & Purity (\%) & Yield (\%) \\ \hline
  traditional &  93.7  & 15.6    \\ 
  $nnet \geq 0.50$ & 91.5 & 23.2  \\
  $nnet \geq 0.60$ & 92.8 & 22.6  \\
  $nnet \geq 0.70$ & 93.6 & 22.1  \\
  $nnet \geq 0.80$ & 95.4 & 21.1  \\
  $nnet \geq 0.90$ & 97.2 & 19.2  \\
  $nnet \geq 0.95$ & 99.0 & 16.2  \\ \hline
  \multicolumn{3}{c} 
{Table 1: Purity and yield factors for Monte Carlo $\Lambda$'s.} 
\end{tabular} 
\end{table}

\begin{figure}[t]
\centerline{\psfig{figure=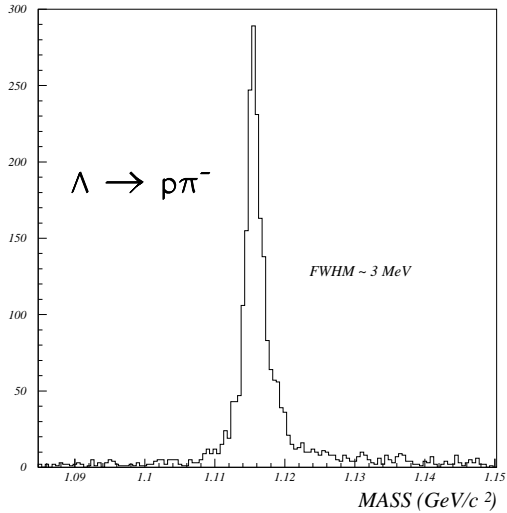,height=3in}}
 \caption  { $\Lambda$ invariant mass spectrum from neural network cut.}
 \label{MLAM_NNET}
\end{figure}

The ultimate test of a neural  network filter
is its performance on actual data.
Although yield and purity
factors cannot be calculated,
the overall performance can be judged by
comparing the invariant mass distributions
which results from cutting on $o$ to
those which result from the traditional method of
cutting. 
The distribution for EOS $\Lambda$'s obtained by
requiring $o \geq 0.95$ is shown
in Fig.~\ref{MLAM_NNET} on the same scale as the distribution
of Fig.~\ref{MLAM_TRAD}. A higher peak and lower background
are clearly evident in the neural net filtered
distribution. When invariant mass cuts are 
applied (1112 MeV/$c^{2}$ $\leq M_{\Lambda} \leq$ 1120 MeV/$c^{2}$),
the neural net method gives 1797 $\Lambda$ candidates
as opposed to 1362 for the traditional method.

\section{Conclusions}

For the EOS TPC data the neural network approach 
results in significant enhancements in both the
yields and purities of $\Lambda$'s and K$^{0}_{\mathrm{s}}$'s
compared to the straightforward method of cutting
in seven dimensional parameter space. 
The $\Lambda$ candidates  in the peaks of
Figs.~\ref{MLAM_TRAD} and \ref{MLAM_NNET} 
were projected onto the seven parameter
axes and a comparison of the resulting 
distributions was made.
In general, the edges of the neural
network filtered events are
less sharp; lending support to the
intuitive notion that the neural filter
finds a smoother hypersurface in the parameter 
space.

When working with neural network filters it is
obviously important to insure that the training
set matches  the data to be filtered as closely
as possible. This was observed in the present 
study when the neural network was initially
trained with a set of ARC events which did
not include the coalescence of protons and
neutrons to form deuterons. The result
were V0 neural networks which performed
only marginally better than the seven
parameter cuts method.

The neural network topology of Fig.~\ref{TOPOLOGY}
was found to work so well that other  
topologies were not investigated. It is possible
that alternative network architectures could
give even better results. One of the disadvantages
of the neural network approach is that there exists
no {\em a priori\/} prescription for evaluating 
various topologies --- one must simply proceed
through trial and error. 

The cpu time
spent in training the networks was not significant;
however, the time spent in generating the training
set was quite large: $\sim$5 cpu minutes/event
$\times$ $\sim$17,000 strange ARC events. Since
the GEANT simulations were being done
anyway in order to study acceptance and efficiency
issues the net cpu overhead on data processing
due to neural network training can be considered
to be negligible.

\ack
The author would like to thank David Kahana for providing
the ARC events.
This work is supported in part by the US Department of Energy under 
contracts/grants DE-AC03-76SF00098, DE-FG02-89ER40531, DE-FG02-88ER40408, 
DE-FG02-88ER40412, DE-FG05-88ER40437, and by the US National Science 
Foundation under grant PHY-9123301.

\end{document}